\documentclass[twocolumn,aps,prl,showpacs,amsmath,amssymb,superscriptaddress,floatfix]{revtex4-2}
\usepackage[breaklinks=true,colorlinks,citecolor=blue,linkcolor=blue,urlcolor=blue]{hyperref}
\usepackage{natbib}
\usepackage[T1]{fontenc}
\usepackage[utf8]{inputenc} 
\usepackage{newunicodechar}
\newunicodechar{−}{\textminus}
\usepackage{lineno}
\usepackage{textcomp}
\usepackage{amsmath}
\usepackage{graphicx}
\usepackage{multirow}
 \usepackage{makecell}
 \usepackage{booktabs}
\usepackage{bbold}
\usepackage{lipsum}
\usepackage{array}
\newcolumntype{P}[1]{>{\centering\arraybackslash}p{#1}}

\def\be{\begin{equation}}
\def\ee{\end{equation}}

\def\bea{\begin{eqnarray}}
\def\eea{\end{eqnarray}}

\makeatletter
\usepackage{lipsum}
\usepackage{epsfig}

\usepackage[colorlinks,linkcolor=blue]{hyperref}
\usepackage[nameinlink,capitalise]{cleveref}

\usepackage{subfigure}
\usepackage{color}
\usepackage{physics}

\definecolor{Red}{rgb}{1,0,0}
\definecolor{Blu}{rgb}{0,0,1}
\definecolor{Green}{rgb}{0,1,0}

\makeatother
\usepackage{amssymb}
\usepackage{babel}

\usepackage{tikz,xcolor,hyperref}
\definecolor{lime}{HTML}{A6CE39}
\DeclareRobustCommand{\orcidicon}{%
	\begin{tikzpicture}
	\draw[lime, fill=lime] (0,0)
	circle [radius=0.16]
	node[white] {{\fontfamily{qag}\selectfont \tiny ID}};
	\draw[white, fill=white] (-0.0625,0.095)
	circle [radius=0.007];
	\end{tikzpicture}
	\hspace{-2mm}
}

\foreach \x in {A, ..., Z}{%
	\expandafter\xdef\csname orcid\x\endcsname{\noexpand\href{https://orcid.org/\csname orcidauthor\x\endcsname}{\noexpand\orcidicon}}
}

\begin{document}

\title{Coupled Electronic-Structure and Phonon-Transport Driven by Bonding Reconstruction in Ba–B–C Clathrates}

\author{Ghulam Hussain\orcidH}
\email{ghussain@szu.edu.cn}
\affiliation{International Research Centre MagTop, Institute of Physics, Polish Academy of Sciences, Aleja Lotnik\'ow 32/46, PL-02668 Warsaw, Poland}
\affiliation{Institute for Advanced Study, Shenzhen University, Shenzhen 518060, China}
\author{Dario Massa\orcidK}
\affiliation{IDEAS Research Institute, 13 Krakowskie Przedmieście Street
00-071 Warsaw, Poland}
\author{Muhammad Usman}
\affiliation{College of Electronics and Information Engineering, Shenzhen University, Shenzhen, 518060, Guangdong, China}
\author{Rajibul Islam\orcidC}
\affiliation{Department of Physics, University of Alabama at Birmingham, Birmingham, Alabama 35294, USA}
\author{Magdalena Birowska\orcidL}
\email{magdalena.birowska@uw.edu.pl}
\affiliation{Institute of Theoretical Physics, Faculty of Physics, University of Warsaw, Warsaw 02-093,
Poland}
\author{Carmine Autieri\orcidA}
\affiliation{International Research Centre MagTop, Institute of Physics, Polish Academy of Sciences, Aleja Lotnik\'ow 32/46, PL-02668 Warsaw, Poland}
\author{Xiaoguang Li\orcidF}
\email{xgli@szu.edu.cn}
\affiliation{Institute for Advanced Study, Shenzhen University, Shenzhen 518060, China}

\begin{abstract}
Composition-driven changes in electron count, structural symmetry and bonding environment can couple the electronic and thermal responses in a material. Here, with density-functional theory and machine-learned interatomic potentials (MLIPs), we compare cubic BaB$_3$C$_3$ with its B$\rightarrow$C substituted tetragonal counterpart, BaB$_2$C$_4$. The additional valence electron and accompanying change in crystal symmetry drive a transition from metallic BaB$_3$C$_3$ to semiconducting BaB$_2$C$_4$ and the formation of short/strong C–C bonds.\ The reconstructed bonding network hardens the phonon spectrum, with the maximum frequency increasing from $\sim$800 to 900 cm$^{-1}$. At room temperature, the in-plane lattice thermal conductivity ($\kappa$) increases from 7.6 to 18.7 Wm$^{-1}$K$^{-1}$, while the thermal transport alters from isotropic to anisotropic ($\kappa$$_x$= $\kappa$$_y$ > $\kappa$$_z$) for BaB$_2$C$_4$. This enhancement is associated with the strengthened bonding network, higher phonon frequencies/group velocities, longer lifetimes and larger mean free paths. Machine learning potentials accurately reproduced the DFT-level energies and forces with negligible errors, and the harmonic and anharmonic features are benchmarked against explicit first-principles calculations. Our findings demonstrate that minimal B$\rightarrow$C substitution provides a viable strategy to link composition- and structure-associated bonding reconstruction with correlated electronic and phononic alterations.
\end{abstract}

\maketitle

Controlling heat transport in solids is crucial for thermal management, energy conversion, and thermal-information technologies \cite{henry2020five,forman2016estimating,wehmeyer2017thermal,alam2023sign,utaka2019application}.\ Heat is mainly transported through electronic, lattice (phononic), and magnetic (magnonic) excitations \cite{singleton2001band,holland1964phonon,qian2021phonon,hanus2021thermal,muhammad2024signature,hess2003magnon,chen2020synthesis,hess2001magnon}.\ In non-metallic crystals, phonons dominate thermal transport. While electronic thermal conductivity is tunable through electrical conductivity (Wiedemann–Franz law), magnetic excitations provide an alternative field-sensitive transport channel \cite{franz1853ueber,kimling2015spin,hess2019heat}.\ In contrast, controlling thermal transport in semiconductors and insulators is challenging because phonons are not directly manipulated by electromagnetic fields \cite{ziman1960electrons}.\

The lattice thermal conductivity is governed by phonon frequencies, group velocities, and phonon lifetimes, which are strongly influenced by atomic mass, crystal symmetry, and the strength of interatomic bonding \cite{hussain2025engineering,raya2026thermal,kizuka2015temperature,lyeo2006thermal,hussain2025phase}.\ Accordingly, thermal transport has been tailored through external perturbations, including electric and magnetic fields, temperature, strain, pressure, and light \cite{zhang2025electric,qin2017external,muhammad2024electronic,seijas2019strain,meng2019thermal,mcguire2018measurements,wan2023polymer,shin2019light}, as well as through chemical and structural modifications \cite{reifenberg2007thickness,lee2013phonon,liu2023low}.\ Among these, chemical substitution can simultaneously modify the electron count, crystal symmetry and the bonding framework, thereby establishing a direct link between electronic reconstruction and phonon thermal transport. B–C clathrates offer a platform to explore this interplay because substantial bonding rearrangements are feasible through small compositional changes \cite{cui2022prediction}.\ However, how compositional-induced changes in electron count and structural symmetry collectively impacts the electronic character, the bonding network and phonon transport remains insufficiently understood.\ Above all, the microscopic origin of the enhanced lattice thermal conductivity due to bonding reconstruction in these compounds has not been established.

Here, we study the coupled electronic and phononic responses of BaB$_3$C$_3$ and its B$\rightarrow$C-substituted counterpart BaB$_2$C$_4$ using DFT, bonding analysis, and machine-learned potentials. The MLIPs reproduce DFT energies and forces with high accuracy to efficiently evaluate the harmonic and anharmonic lattice dynamics. Our findings establish that replacing one B atom with C increases the electron count by one per formula unit and reduces the crystal symmetry from cubic to tetragonal. This transforms metallic BaB$_3$C$_3$ into the semiconducting BaB$_2$C$_4$ compound, which exhibits an indirect bandgap of 0.33 eV and features strong C-C bonds. We further demonstrate that the reconstructed bonding environment hardens the phonon spectrum, while concomitantly increasing the room-temperature $\kappa$ from 7.6 to 18.7 W m$^{-1}$ K$^{-1}$. This enhancement is driven by microscopic contributions such as improved phonon frequencies, faster group velocities, longer phonon lifetimes, and larger mean free paths. 

\begin{figure*}[t]
	\centering
	\includegraphics[scale=0.25, angle=0] {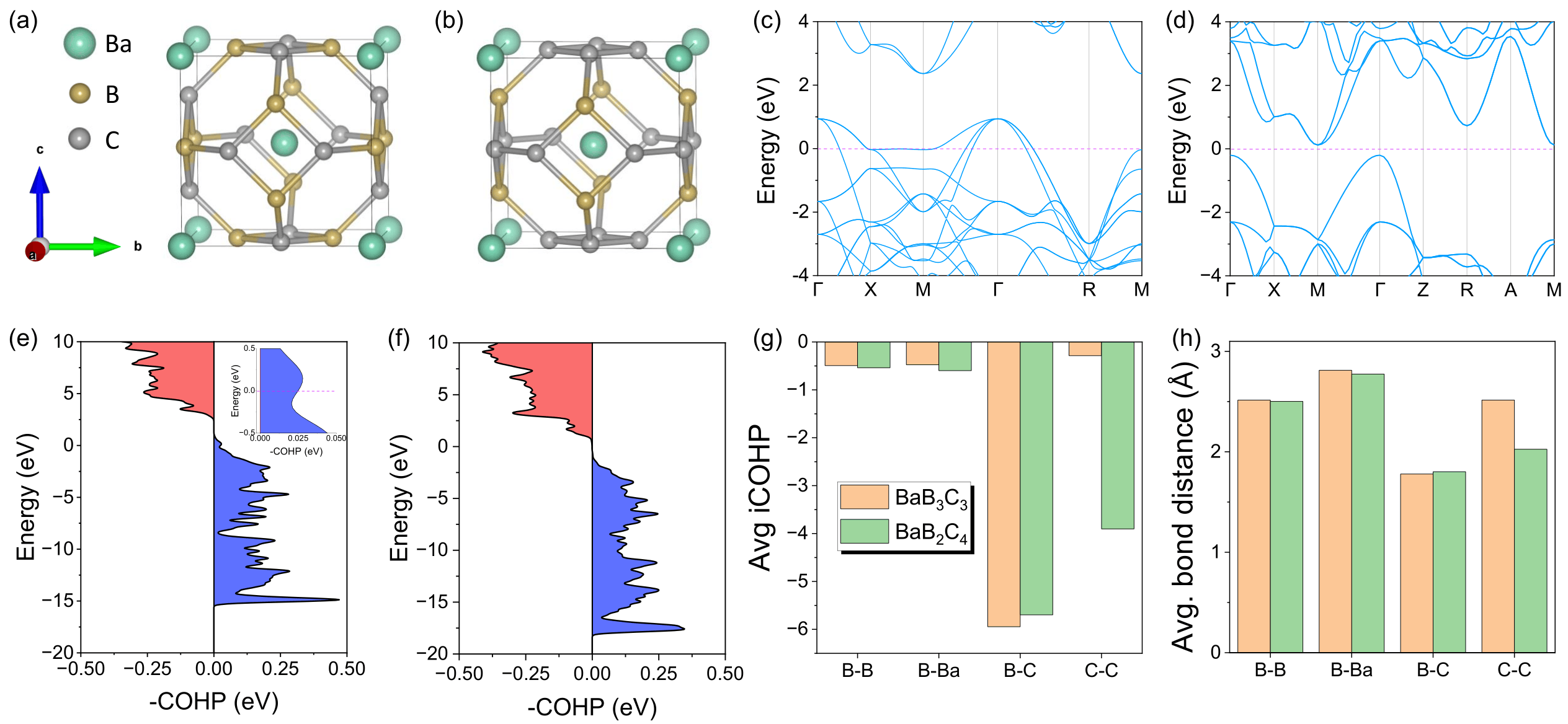}
	\vspace{-1em}
    \caption{Crystal structures, (a) Cubic unit cell of BaB$_3$C$_3$, with Ba, B, and C atoms shown as teal, gold, and grey spheres, respectively. (b) Tetragonal unit cell of BaB$_2$C$_4$ obtained by substituting one B per formula unit by C. Electronic bandstructures calculated for (c) BaB$_3$C$_3$ revealing metallic character, while (d) BaB$_2$C$_4$ displays semiconducting behavior. Crystal orbital Hamilton population ($-\mathrm{COHP}$) of (e) BaB$_3$C$_3$  and (f) BaB$_2$C$_4$; the inset of (e) shows $-\mathrm{COHP}$ in the vicinity of Fermi level, highlighting the presence of unpopulated bonding states above E$_F$, giving rise to the characteristic metallic behavior. Upon B$\rightarrow$C substitution, the bonding manifold terminates and the $-\mathrm{COHP}$ signal drops to zero. (g) Average integrated COHP (iCOHP) and (h) average bond distances for some selective pairs of atoms like B–B, B–Ba, B–C, and C–C in BaB$_3$C$_3$ (orange) and BaB$_2$C$_4$ (green). B$\rightarrow$C substitution markedly strengthens the C–C bonds.}
    \label{1}
\end{figure*}

We employed Density-functional theory (DFT) to perform calculations using VASP \cite{kresse1993ab,kresse1996efficiency} with the projector augmented-wave method and a plane-wave cutoff of 550 eV \cite{kresse1999ultrasoft}. Structural optimizations employed a $10\times10\times10$ k-point mesh, while electronic structures were calculated using the HSE06 hybrid functional \cite{heyd2003hybrid,kecik2016optical}.\ Chemical bonding analysis was carried out using LOBSTER through projected and integrated crystal orbital Hamilton populations (COHP/iCOHP) \cite{maintz2016lobster}.\ Ab initio molecular-dynamics simulations were acquired at 300 K using 112-atom supercells with the NVT ensemble (6000 steps with a 2 fs time step)  \cite{nose1984unified}.\ The resulting structural configurations, their energies, and forces were used to train MACE interatomic potentials \cite{batatia2022mace}.\ Second- and third-order interatomic force constants (IFCs) were generated using Phonopy and Phono3py, respectively, employing a $2\times2\times2$ supercell and a 6 \AA\ interaction cutoff \cite{togo2023first}.\ $\kappa$ was calculated by solving the full linearized Boltzmann transport equation over 100–500 K. The MACE-based results for BaB$_3$C$_3$ were then benchmarked via independent DFT calculations by employing Quantum ESPRESSO \cite{giannozzi2009quantum} and ShengBTE \cite{li2014shengbte}.\ The full computational details, including convergence tests and MACE training parameters, are provided in the Supplementary Material \cite{link}.

\begin{figure*}[t]
	\centering
	\includegraphics[scale=0.25, angle=0] {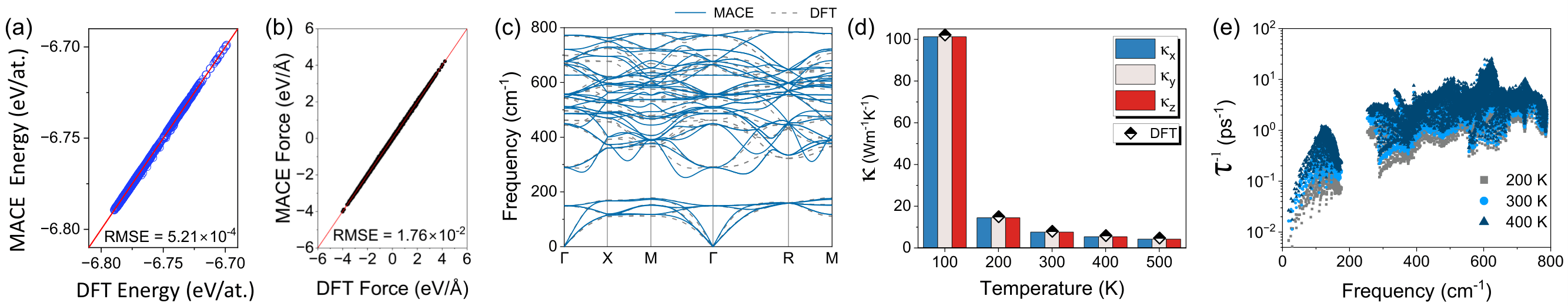}
    \vspace{-1em}
	\caption{(a, b) MACE energies and forces versus the DFT reference values for the BaB$_3$C$_3$ compound, with RMSEs of 5.21$\times$10$^{-4}$ eV/atom and 1.76$\times$10$^{-2}$ eV/\AA, respectively. The red dashed lines reveal the ideal MACE–DFT agreement validating the MACE MLIPs. Lattice dynamics and phonon thermal transport of cubic BaB$_3$C$_3$ compound: (c) Phonon dispersions along the high-symmetry path, where solid Blue lines correspond to MACE frequencies and dashed Grey lines represent DFT based phonon spectrum. (d) Lattice thermal conductivity ($\kappa$) as a function of temperature obtained along the three Cartesian directions; the colored bars represent MACE calculations showing isotropic transport ($\kappa$$_x$= $\kappa$$_y$= $\kappa$$_z$), while the diamonds represent reference DFT values. (e) Phonon scattering rates as a function of frequency calculated using MACE at different temperatures (200, 300, and 400 K); which increase with temperature, in agreement with Umklapp‑dominated three‑phonon scattering.}
    \vspace{-0.5em}
    \label{2}
\end{figure*}

Fig.~\ref{1} illustrates the crystal structures, electronic bandstructures, and chemical-bonding characteristics of pristine BaB$_3$C$_3$ and the corresponding B$\rightarrow$C substituted tetragonal BaB$_2$C$_4$.\ The relaxed structure for cubic BaB$_3$C$_3$ with space group $Pm\bar{3}n$ and lattice parameter a = 5.030 \AA, is shown in Fig.~\ref{1}(a) retaining the high-symmetry, whereas replacing one B with one C atom per formula unit lowers the symmetry and transforms it into the tetragonal BaB$_2$C$_4$ (space group I4/mmm, and lattice parameters, a = 4.950 \AA, and c = 5.003 \AA) structure as displayed in Fig.~\ref{1}(b); the values of the lattice constants are in agreement with the literature \cite{di2022high,cui2022prediction}.\ Despite the seemingly trivial compositional change, the substitution leads to a striking reconstruction of both the electronic structure and the local bonding environment. The electronic bandstructure shown in Fig.~\ref{1}(c) reveals the metallic character in cubic BaB$_3$C$_3$ compound as the bands cross the Fermi level (E$_F$). In contrast, the B$\rightarrow$C-substituted tetragonal BaB$_2$C$_4$ system shifts the E$_F$, separating the valence- and conduction-band edges. This behavior leads to a metal-to-semiconductor transition with an indirect narrow bandgap of approximately 0.33 eV (see Fig.~\ref{1}(d)). The electronic reconstruction is not merely a rigid shift of the E$_F$; it is associated by a distinct reorganization of chemical bonding environment. 

To elucidate the bonding reconstruction, we analyzed the COHP, with the corresponding $-\mathrm{COHP}$ distributions demonstrated in Fig.~\ref{1}(e, f).\ The positive (negative) values denote bonding (antibonding) states; the occupied states are dominated by bonding, while antibonding contributions become relevant towards higher energies. The inset of Fig.~\ref{1}(e) shows $-\mathrm{COHP}$ in the vicinity of E$_F$, indicating the presence of bonding states above E$_F$, underscoring the fact that a fraction of the bonding manifold remains unoccupied.\ After B$\rightarrow$C substitution, the electron count increases by one valence electron per formula unit thereby filling the unoccupied bonding manifold and thus causing the upward shift of E$_F$.\ The $-\mathrm{COHP}$ signal drops to zero vanishing the bonding manifold at E$_F$.\ Comparison of the two crystal systems reveals a redistribution of bonding interactions following substitution, conforming with the changes in the electronic bandstructures.\ Fig.~\ref{1}(g) provides a quantitative measure of this reconstruction for some specific bonds, where the average integrated COHP (iCOHP) values for BaB$_3$C$_3$ are approximately −0.49 eV for B–B, −0.48 eV for B-Ba, −5.95 eV for B–C, and −0.29 eV for C–C interactions. On the other hand, the corresponding values for BaB$_2$C$_4$ are around −0.54, −0.60, −5.70, and −3.90 eV, respectively. The B–B and B–Ba interactions become relatively stronger, as signified by their more negative iCOHP values, whereas the B–C interactions change only modestly. Most strikingly, the contribution of C–C interaction increases substantially in magnitude in tetragonal BaB$_2$C$_4$ compound.\ This analysis is further supported by the magnitudes of average bond distances in Fig.~\ref{1}(h).\ The average B–C bond distance slightly changes from $\sim$1.78 \AA (BaB$_3$C$_3$) to 1.80 \AA (BaB$_2$C$_4$), whereas B–Ba and B–B bonds undergo smaller decrease in magnitudes.\ In contrast, the BaB$_2$C$_4$ markedly shows shorter C–C interactions (1.95 \AA) compared with the longer C–C bond distance (2.55 \AA) in cubic BaB$_3$C$_3$ designating the formation of stronger covalent C–C interactions.\ This bonding reconstruction supplies the microscopic basis for the subsequent alterations in lattice dynamics, and phonon thermal transport in BaB$_2$C$_4$.

Because the bonding reconstruction and symmetry change can directly influence the harmonic and anharmonic interatomic force constants, accurate modeling of the potential energy surface is crucial to model lattice thermal conductivity.\ To achieve first-principles accuracy at lower computational cost, here MACE-based \cite{batatia2022mace} MLIPs are trained on reference data generated through AIMD simulations \cite{link}. The quality of the trained MLIPs was determined by comparing the MACE predicted energies and forces with the DFT reference values on the validation set as manifested in Fig.~\ref{2}(a, b). The calculated energies and forces exhibit essentially linear correlations following the one-to-one relationship as indicated by red line revealing the ideal MACE–DFT agreement for these clathrate materials. The MACE-energy and MACE-force root-mean-square errors (RMSEs) for BaB$_3$C$_3$ are 5.21$\times$10$^{-4}$ eV/atom and 1.76$\times$10$^{-2}$ eV/\AA, respectively. The small RMSE values demonstrate that the trained MLIP reproduces the DFT energies and forces with high fidelity over the sampled configurations. This level of consistency provides a reliable basis for using the MACE MLIPs in subsequent calculation of lattice dynamics and phonon thermal transport.

The resulting trained MACE model was then used for computing the phonon spectra and lattice thermal transport properties.\ Fig.~\ref{2}(c) presents the phonon bandstructure of cubic BaB$_3$C$_3$ compound obtained using the MACE MLIPs and DFT, respectively.\ The calculated phonon dispersion shows no imaginary frequencies, confirming the dynamical stability of this cubic structure.\ The spectrum extends to $\sim$800 cm$^{-1}$ and the acoustic branches exhibit pronounced dispersion near the $\Gamma$ point, identifying these modes as the principal channels for heat transport.\ The temperature dependence of the calculated $\kappa$ is shown in Fig.~\ref{2}(d), which reveals isotropic phonon thermal transport at all temperatures.\ A systematic decrease in the thermal conductivity can be observed with rising temperature that is attributed to increased phonon–phonon scatterings as illustrated in Fig.~\ref{2}(e). The calculated phonon scattering rates (\(\tau ^{-1}\)) display noticeable frequency and temperature dependent features. Below 180 \(\text{cm}^{-1}\), the acoustic phonon-phonon scatterings scale sharply with frequency, whereas the phononic gap in the scattering spectrum corresponds to a range of frequency (180-250 \(\text{cm}^{-1}\)) with almost no allowed phonon interactions. The rising temperature increases the phonon population, which consequently enhances the probability of phonon-phonon interactions. The temperature dependent \(\tau ^{-1}\) is significantly influenced by the Umklapp processes, which become dominant at higher temperatures \cite{cui2023phononic}. 

The above MACE-based findings establish cubic BaB$_3$C$_3$ as a dynamically stable and isotropic heat conducting material with a room-temperature $\kappa$ of 7.6 Wm$^{-1}$K$^{-1}$. To validate the accuracy of these MACE-MLIPs, we benchmarked its performance against explicit DFT calculations.\ First, we scrutinize the harmonic lattice dynamics by comparing phonon bandstructures along the high-symmetry paths (see Grey curve in Fig.~\ref{2}(c)). The low-frequency acoustic phonon modes obtained from the MACE IFCs exhibit excellent agreement with the full-DFT calculation, while the high frequency optical branches show some minor local deviations-confirming that the MLIPs derived from MACE framework accurately captures the interatomic force fields.\ Importantly, the MACE-based $\kappa$ values (shown by bars) match the explicit DFT calculations (shown by diamond symbols) almost perfectly at all temperatures (Fig.~\ref{2}(d)).\ This rigorous consistency validates the capability of the MACE-based MLIPs to predict both harmonic and anharmonic characteristics in BaB$_3$C$_3$ at a fraction of the computational cost.

\begin{figure}[h!]
\centering
\includegraphics[width=0.99 \linewidth]{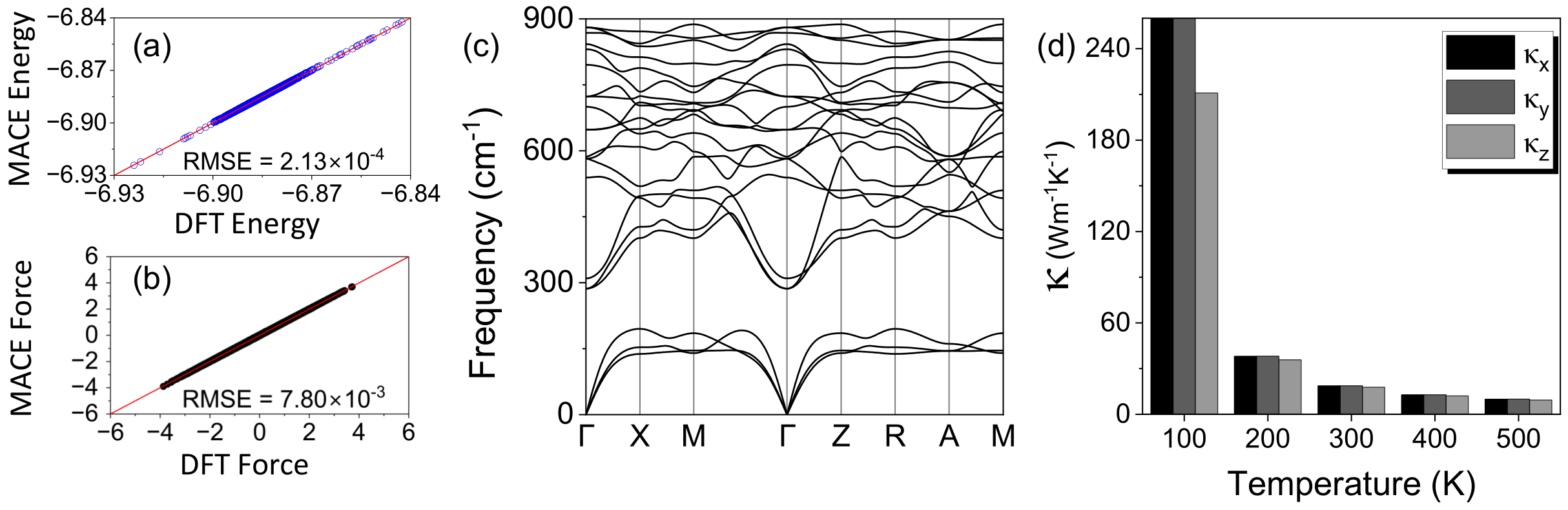}
\vspace{-2em}
\caption{(a, b) MACE energies and forces versus the DFT reference values for BaB$_2$C$_4$ compounds, with RMSEs of 2.13$\times$10$^{-4}$ eV/atom and 7.80$\times$10$^{-3}$ eV/\AA, respectively.\ The red dashed line reveals the ideal MACE–DFT agreement validating the MACE MLIPs. (c) Phonon bandstructure of BaB$_2$C$_4$ showing a hardened phonon spectrum (frequency reaching $\sim$900 cm$^{-1}$),\ reflecting enhanced effective IFCs associated with the reconstructed bonding network, consistent with the stronger bonding interactions identified by LOBSTER. (d) Lattice thermal conductivity of BaB$_2$C$_4$; indicating anisotropic transport ($\kappa_x \approx \kappa_y > \kappa_z$), as required by tetragonal symmetry.\ The B$\rightarrow$C substitution enhances in-plane $\kappa$ by a factor of $\sim$2.5 compared to the cubic phase.}
\vspace{-0.9em}
\label{3}
\end{figure}

After establishing the validity of the MACE architecture, we analyzed the phonon dispersion and temperature-dependent $\kappa$ of tetragonal BaB$_2$C$_4$ using the MACE interatomic potentials.\ The MACE-calculated energies and forces shown in Fig.~\ref{3}(a, b) exhibit essentially high accuracy and fidelity with very small RMSEs (2.13$\times$10$^{-4}$ eV/atom for energy and 7.80$\times$10$^{-3}$ eV/\AA\ for force).\ The phonon bandstructure in Fig.~\ref{3}(c) contains no imaginary branches, confirming the dynamical stability of the substituted tetragonal crystal. In comparison with cubic BaB$_3$C$_3$, we observed substantially hardened phonon spectrum, with the optical phonon branches extending to approximately 900 cm$^{-1}$. The upward shift in the frequency is consistent with the reconstructed covalent bonding induced by the substitution of B with C atom, particularly, the formation of short and strong C–C interactions as identified in the LOBSTER analysis.\ On the other hand, the acoustic branches remain dispersive near the $\Gamma$-point providing the principal low-frequency channels for phonon transport.\ Moreover, the lattice thermal conductivity as a function of temperature for the tetragonal structure is shown in Fig.~\ref{3}(d), which shows anisotropic phonon transport ($\kappa_x \approx \kappa_y > \kappa_z$) and a pronounced decrease in $\kappa$ with increasing temperature.\ The in-plane lattice thermal conductivity ($\kappa_x , \kappa_y$) reaches around 18.7 Wm$^{-1}$K$^{-1}$ at room temperature, substantially exceeding the value (7.6 Wm$^{-1}$K$^{-1}$) obtained for BaB$_3$C$_3$ compound. This marked increase in $\kappa$ results from the combined effects of modified harmonic and anharmonic lattice dynamics. 

\begin{figure}[h!]
\centering
\includegraphics[width=1.0 \linewidth]{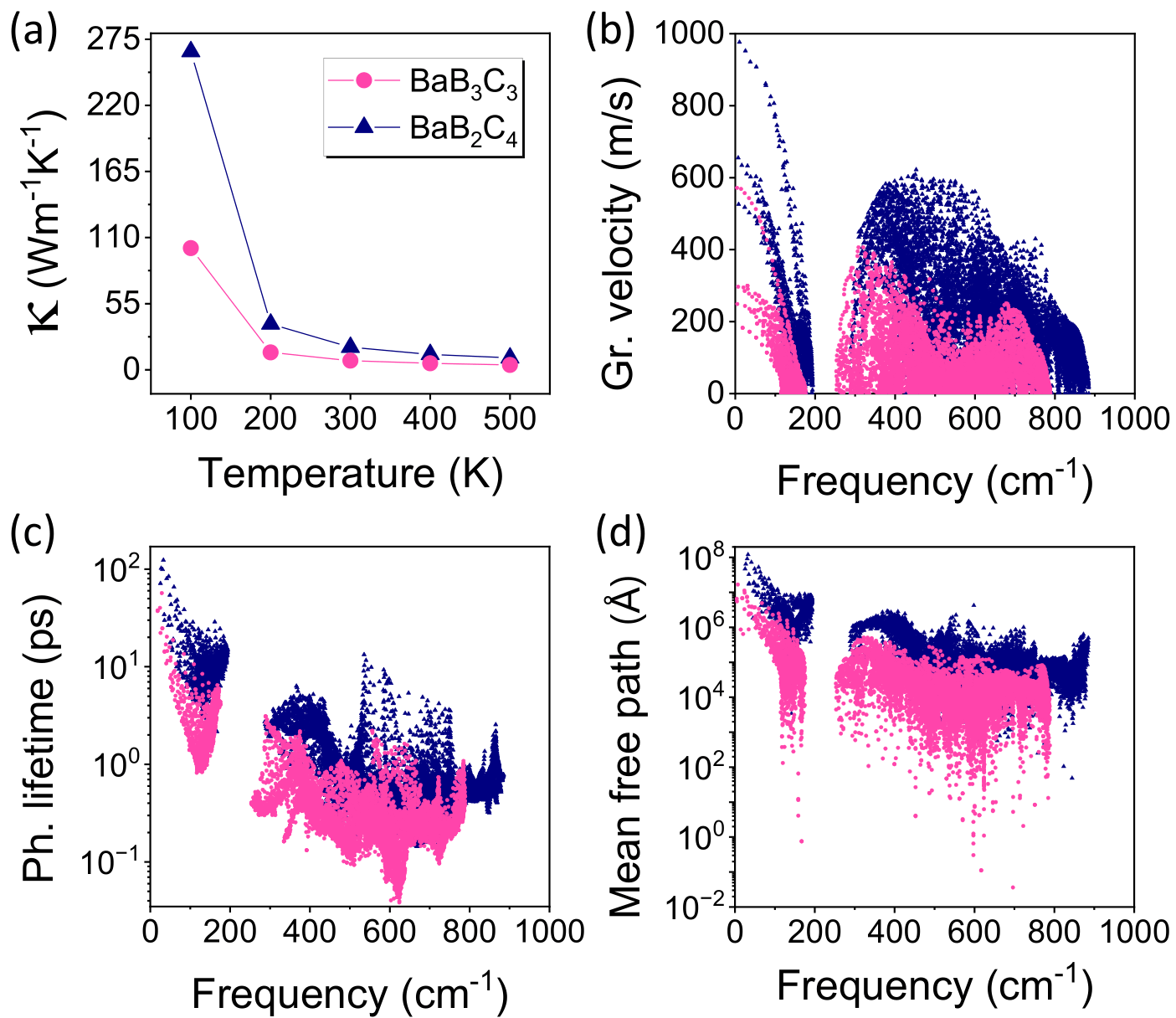}
\vspace{-1.7em}
\caption{Comparative analysis of (a) lattice thermal conductivity $\kappa$, (b) phonon group velocities, and (c) corresponding phonon lifetimes for BaB$_3$C$_3$ (represented by pink spheres) and BaB$_2$C$_4$ (denoted by blue rectangles) compounds, respectively.\ The substantially faster phonon group velocities and longer phonon lifetimes yield (d) larger mean free paths in tetragonal BaB$_2$C$_4$, contributing to its enhanced $\kappa$ relative to BaB$_3$C$_3$.}
\label{4}
\end{figure}

To investigate the microscopic origin of the enhanced $\kappa$ following B$\rightarrow$C substitution, we present a direct comparison of the phonon thermal transport of BaB$_3$C$_3$ and BaB$_2$C$_4$ compounds.\ As shown in Fig.~\ref{4} (a), BaB$_2$C$_4$ reveals a consistently higher $\kappa$ at all temperatures, indicating that its enhanced heat transport is intrinsic to the modified lattice dynamics rather than being limited to a particular narrow temperature range. Notably, the in-plane thermal conductivity reaches approximately 18.7 Wm$^{-1}$K$^{-1}$, compared with 7.6 Wm$^{-1}$K$^{-1}$ for BaB$_3$C$_3$, highlighting an enhancement of about $\sim$2.5 times. The phonon group velocities in Fig.~\ref{4} (b) manifest the first microscopic contribution to this enhancement. Across a wide frequency range, we observed larger phonon group velocities for BaB$_2$C$_4$ compound than those of BaB$_3$C$_3$. This difference is particularly reflected in the low- and intermediate-frequency modes, which are expected to make significant contributions to phonon heat transport. These larger group velocities are in agreement with the hardened phonon spectrum of BaB$_2$C$_4$ in Fig.~\ref{3}(a), and indicative of more efficient thermal transport through the reconstructed covalent framework. The second major contribution arises from the anharmonic phonon-phonon interactions as exhibited from the phonon lifetimes in Fig.~\ref{4}(c).\ Longer lifetimes for BaB$_2$C$_4$ structure imply fewer scattering events allowing the propagation of heat-carrying phonons over longer distances before being scattered. Thus, the increased thermal transport in BaB$_2$C$_4$ compound cannot be attributed solely to the upward shift in phonon frequencies; it originates from the combined modification of phonon spectra and anharmonic interactions. The combined impact of both phonon group velocities and the relaxation times is directly dictated by the mean free path as shown in Fig.~\ref{4}(d).\ As a direct consequence of the enhanced phonon group velocities and longer lifetimes (\(\Lambda = v \cdot \tau\)), the BaB$_2$C$_4$ reveals longer mean free path compared to BaB\({}_{3}\)C\({}_{3}\) compound. In the acoustic region, this elongation in mean free path is exceptionally pronounced, where the values reach to $\sim$10$^{8}$ \AA, thus allowing the critical modes to travel much longer displacements before they scatter. As a result, such extended propagation pathways provide definitive unified microscopic evidence that explain the $\sim$2.5-fold enhancement of the in-plane $\kappa$. The comparison presented here showcases the implication of B$\rightarrow$C substitution on the harmonic and anharmonic features governing the phonon thermal transport. 

In summary, we have illustrated that a minimal chemical modification of B-C clathrate can produce pronounced and correlated alterations in its electronic structure, interatomic bonding, lattice dynamics, and phonon thermal transport. Starting from cubic BaB$_3$C$_3$, the single B$\rightarrow$C substitution transmutes it into tetragonal BaB$_2$C$_4$, reduces the crystal symmetry, increases the electron count and simultaneously reconstructs the bonding network. This seemingly small compositional modification induces metal-to-semiconductor transition and substantially strengthens and shortens the C–C bonds. In addition, the lattice dynamics are significantly modified; the phonon spectrum of BaB$_2$C$_4$ compound hardens, anisotropy is induced in $\kappa$ due to the reduced symmetry, larger phonon group velocities, longer phonon lifetimes and mean free paths are observed. These factors appreciably enhance the phonon heat transport; particularly the room-temperature $\kappa$ from 7.6 to 18.7 Wm$^{-1}$K$^{-1}$. Importantly, the MACE MLIPs reproduced the DFT energies and forces accurately, reliably capturing the harmonic phonon lattice dynamics and anharmonic interactions, as validated against DFT calculations. Overall, these findings demonstrate that targeted chemical substitution provides a viable approach to systematically reshape the coupled electronic structure, bonding, lattice-dynamics, and thermal transport in B–C clathrate frameworks. 

\vspace{0.5em}
\begin{acknowledgments}
This work was supported by the National Natural Science Foundation of China Grants No. 12274301 and W2433014. The research was also supported by the Key Project of Department of Education of Guangdong Province Grant No. 2024ZDZX1014. C. A. was supported by the Foundation for Polish Science project “MagTop” No. FENG.02.01-IP.05-0028/23 co-financed by European Union from funds of Priority 2 of European Funds for Smart Economy Program 2021–2027 (FENG).
\end{acknowledgments}

\bibliographystyle{apsrev4-2}
\vspace{-1.5em}
\bibliography{ref}

\end{document}